\documentclass[aps,prd,twocolumn,showpacs,showkeys,superscriptaddress,floatfix,elsart]{revtex4-1}
\usepackage{latexsym}
\usepackage{amssymb}
\usepackage{amsmath}
\usepackage{amscd}
\usepackage{amsthm}
\usepackage{graphicx}
\usepackage{textcomp}
\usepackage{colortbl}
\usepackage{hyperref}
\usepackage[font={footnotesize,it}]{caption}
\usepackage{multirow}
\usepackage[T1]{fontenc}
\usepackage{ae,aecompl}
\begin{document}
\title{Lyra's cosmology of homogeneous and isotropic universe in Brans-Dicke theory}

\author{Rajendra Prasad}
\email[]{drrpnishad@gmail.com}
\affiliation{Department of Physics, Galgotias College of Engineering and Technology, Greater Noida 201310, India}
\author{Lalit Kumar Gupta}
\email[]{lalitl247@gmail.com}
\affiliation{Department of Physics, Krishna Engineering College, Ghaziabad, India}
\author{Anil Kumar Yadav}
\email[]{abanilyadav@yahoo.co.in}
\affiliation{Department of Physics, United College of Engineering and Research,Greater Noida - 201310, India}
\begin{abstract}
In this paper, we investigate a scalar field Brans-Dicke cosmological model in Lyra's geometry which is based on the modifications in geometrical term as well as energy term of Einstein's field equations. We have examined the validity of proposed cosmological model on observational scale by performing statistical analysis from latest $H(z)$ and SN Ia observational data. We find that the estimated values of Hubble's constant and matter energy density parameter are in agreement with their corresponding values, obtained from recent observations of WMAP and Plank collaboration. We also derived deceleration parameter, age of the universe and jerk parameter in terms of red-shift and computed its present values. The dynamics of deceleration parameter in derived model of the universe shows a signature flipping from positive to negative value and also indicates that the present universe is in accelerating phase.\\


\end{abstract}
\maketitle
\section{Introduction}
\label{sec:intro}
In 1915, Einstein had proposed General Theory of Relativity (GTR) and beautifully described the geometry of space and time in elegance with gravity. For recent overviews Generak Relativity and the challenges it faces, see, e. g., \cite{Iorio/2015,Debono/2016,Vishwakarma/2016,Jimenez/2019} and references therein. In this theory, it has been proposed that the energy-momentum tensor appears due to curvature of space and time through famous Einstein's field equation: $R_{ij}-\frac{1}{2}Rg_{ij} = \frac{8\pi G}{c^{4}}T_{ij}$, where $R_{ij}$, $T_{ij}$ are the Ricci curvature tensor and energy momentum tensor respectively. G and c are the Newtonian constant of gravitation and the speed of light in vacuum, and that the indexes $i$, $j$ run from 0 to 3. This equation specifies how the geometry of space and time is influenced by whatever matter and radiation are present, and from the core of Einstein's general theory of relativity. In the literature, various cosmological models have been studied in GTR in different physical contexts \cite{Kumar/2011,Sami/2006,Yadav/2011,Yadav/2011a,Kumar/2011a,Yadav/2011b,Yadav/2012epjp}. Among these models, $\Lambda$CDM model is most successful cosmological model to describe the current features of observed universe but it suffers cosmological constant problem or vacuum catastrophe. The cosmological constant problem gave a clue to think about the modification in GTR. In the literature, various modification in Einstein's theory have been proposed by cosmologist from time to time since its inception \cite{Clifton/2012,Capozziello/2015,
Martino/2015}. In this paper, we have applied modification in both the geometrical term as well as in the energy-momentum term by taking into account Lyra's geometry \cite{Lyra/1951} and Brans-Dicke theory of gravitation \cite{Brans/1961} respectively.\\

Lyra's geometry represents the modification of Riemannian geometry by introducing gauge function into structureless manifold. In this approach, gauge function naturally replace the cosmological constant term and hence the cosmological constant problem. This means that time like constant gauge function or constant displacement field plays the role of cosmological constant - a physically accepted candidate of dark energy which is required to accelerate the universe in its present epoch \cite{Halford/1972}. It is worthwhile to note that the singularity free cosmological models have been investigated in the framework of emergent universe scenario by assuming displacement field as time dependent \cite{Darabi/2015} while Ellis \& Maartens\cite{Ellis/2004} have investigated a cosmological model with normal
matter and a scalar field sources in general relativity. Recently there is an upsurge of interest in scalar fields in general relativity and alternative theories of gravitation in the context of inflationary cosmology. In Singh and Desikan \cite{Singh/1997pramana}, it has been investigated that time varying displacement vector $\beta(t)$ is decreasing function of time. Therefore, the study of cosmological models in Lyra's geometry may be relevant for inflationary models. Some important applications of constant and time varying displacement field in Lyra's geometry are given in Refs. \cite{Pradhan/2005,Yadav/2009,Singh/1992,Singh/1993,Yadav/2011d,Rahaman/2005,Yadav/2010e,Yadav/2018,Yadav/2020prd}.\\

Many experimental and theoretical tests of GTR confirm that the local motion of particle does not affect due to large scale matter distribution that is why Mach's principle could be violated in GTR \cite{Mukherjee/2019}. 
So, Brans and Dicke \cite{Brans/1961} had proposed a modified theory of gravity which was simply formulated to validate Mach's principle. The proposed Brans-Dicke (BD) theory of gravitation not only validates Mach's principle but also describes the dynamics of the universe from inflation era to present accelerating epoch \cite{Maurya/2019,Goswami/2017,Yadav/2019a}. Note that BD theory would also modify the average expansion rate of universe due to appearance of BD coupling constant $\omega$ \cite{Sen/2001}. Recently, Akarsu et al. \cite{Akarsu/2019} have studied dynamical behavior of effective dark energy and the red-shift dependency of the expansion anisotropy in the framework of BD theory. This study reveals that high positive values of $\omega$ imply minimal deviation from the $\Lambda$CDM model while small values of $\omega$ produce huge deviation from standard $\Lambda$CDM model. In fact, BD theory of gravity is an interesting alternative to GTR and effectively introduces a scalar field $\phi$, in addition to the metric tensor field $g_{ij}$. The scalar field $\phi$ plays the role of $G^{-1}$ and BD coupling parameter is constrained as $\omega \simeq 40000$ for its consistency with solar system bounds \cite{Bertotti/2003,Felice/2006}. Several investigations have been carried out in BD cosmology with non-minimal coupling of scalar field \cite{Jawed/2015} and minimally interacting holographic dark energy models \cite{Dasu/2018,Kiran/2015,Ramesh/2016}. Recently, Yousaf \cite{Yousaf/2019a,Yousaf/2019b} has studied some dynamical properties of compact objects within the framework of modified gravity corrections.\\

Motivated by the above discussion, in this paper, we investigate a scalar field Brans-Dicke cosmological model in Lyra's manifold. The outlines of this paper is as follows: Section I is introductory in nature. In section \ref{2}, we have described the basic mathematical formalism of scalar field Brans-Dicke universe in Lyra's manifold. Section \ref{3} deals the statistical analysis of derived model with observational data and estimation of model parameters. In sections \ref{4}, we have computed the present values of deceleration parameter, age of universe and jerk parameter of the model under consideration. The last section is devoted to conclusions.
\section{The basic mathematical formalism of scalar field Brans-Dicke universe in Lyra's manifold}\label{2}
Brans-Dicke field equations in Lyra's manifold is read \cite{Maurya/2019} as
\[
G_{ij}+\frac{3}{2}\psi_{i}\psi_{j}-\frac{3}{4}g_{ij}\psi_{k}\psi^{k} =
\]
\begin{equation}
\label{lm}
 -\frac{8\pi T_{ij}}{\phi c^{4}} -\frac{\omega}{\phi^{2}}\left(\phi_{,i}\phi_{,j}-\frac{1}{2}g_{ij}\phi_{,k}\phi^{,k}\right) -\frac{1}{\phi}(\phi_{,i,j}-g_{ij}\square \phi)
\end{equation}
\begin{equation}
\label{lm-1}
\square \phi = \phi^{,i}_{,i} = \frac{8\pi T}{(3+2\omega)c^{2}}
\end{equation}
where $G_{ij}$, $\psi^{i}$, $\omega$ and $\phi$ are the curvature tensor, displacement vector field of Lyra's geometry, Brans-Dicke coupling constant and Brans-Dike scalar field respectively and the other symbols have their usual meaning in Riemannian geometry. We also suppose that $\psi_{i} = (\beta,0,0,0)$ where $\beta$ is the time like constant displacement vector.\\ 

The isotropic and homogeneous space-time is given by
\begin{equation}
\label{metric}
ds^{2} = c^{2}dt^{2} - a^{2}(t)(dx^{2}+dy^{2}+dz^{2})
\end{equation}
where $a(t)$ is the scale factor which defines the rate of expansion of the universe.\\
The energy momentum tensor of perfect fluid is read as
\begin{equation}
\label{em}
T_{ij} = (\rho +p)u_{i}u_{j} - pg_{ij}
\end{equation}
where $u_{i}u^{i} = 1$ is co-moving co-ordinate.\\

For space-time (\ref{metric}), solving equations (\ref{lm}), (\ref{lm-1}) and (\ref{em}), we obtain the following system of equations
\begin{equation}
\label{fe-1}
3\frac{\dot{a}^{2}}{a^{2}}-\frac{3}{4}\beta^{2} = \frac{8\pi \rho}{\phi c^{2}}+\frac{\omega}{2}\left(\frac{\dot{\phi}}{\phi}\right)^{2}-3\frac{\dot{a}\dot{\phi}}{a\phi}
\end{equation}
\begin{equation}
\label{fe-2}
-2\frac{\ddot{a}}{a}-\frac{\dot{a}^{2}}{a^{2}} - \frac{3}{4}\beta^{2} = \frac{8\pi p}{\phi c^{2}}+\frac{\omega\dot{\phi}^{2}}{2\phi^{2}}+2\frac{\dot{a}\dot{\phi}}{a\phi}+\frac{\ddot{\phi}}{\phi}
\end{equation} 
\begin{equation}
\label{fe-3}
\frac{\ddot{\phi}}{\phi}+3\frac{\dot{a}\dot{\phi}}{a\phi} = \frac{8\pi(\rho-3p)}{(2\omega+3)\phi c^{2}}
\end{equation}
Here, over dot stands for derivative with respect to time.\\
The Hubble parameter is defined as
$H = \frac{\dot{a}}{a}$ $\Rightarrow$ $\frac{\ddot{a}}{a} = \dot{H}+H^{2}$\\
Putting these values in equation (\ref{fe-2}), we obtain\\
\begin{equation}
\label{fe-H}
-2\dot{H} -3H^{2}- \frac{3}{4}\beta^{2} = \frac{8\pi p}{\phi c^{2}}+\frac{\omega\dot{\phi}^{2}}{2\phi^{2}}+2\frac{\dot{a}\dot{\phi}}{a\phi}+\frac{\ddot{\phi}}{\phi}
\end{equation}
The deceleration parameter in term of $H$ is read as\\
$q=-\frac{a\ddot{a}}{\dot{a}^{2}} = -1-\frac{\dot{H}}{H^{2}}$\\
$\dot{H} = -(1+q)H^{2}$\\
From equation (\ref{fe-H}), we obtian\\
\begin{equation}
\label{fe-q}
(2q-1)H^{2} = \frac{8\pi p}{\phi c^{2}}+\frac{\omega\dot{\phi}^{2}}{2\phi^{2}}+2\frac{\dot{a}\dot{\phi}}{a\phi}+\frac{\ddot{\phi}}{\phi}+ \frac{3}{4}\beta^{2}
\end{equation}
Equation (\ref{fe-1}) in term of H is given by
\begin{equation}
\label{fe-1H}
3H^{2}-\frac{3}{4}\beta^{2} = \frac{8\pi \rho}{\phi c^{2}}+\frac{\omega}{2}\left(\frac{\dot{\phi}}{\phi}\right)^{2}-3\frac{\dot{a}\dot{\phi}}{a\phi}
\end{equation}
The continuity equation for perfect fluid is given by
\begin{equation}
\label{ec}
\dot{\rho}+3(\rho + p)H = 0
\end{equation} 
It is easy to obtain that the main equations of the model in standard BD cosmology by introducing two effective parameters as
$$\rho_{eff} = \rho+\frac{3\phi c^{2}}{32 \pi}\beta^{2}$$ \& $$p_{eff} = p+\frac{3\phi c^{2}}{32 \pi}\beta^{2}$$ 
Thus, equations (\ref{fe-1}) and (\ref{fe-2}) recast as
\begin{equation}
\label{eff-1}
3\frac{\dot{a}^{2}}{a^{2}} = \frac{8\pi}{\phi c^{2}}\rho_{eff}+\frac{\omega}{2}\left(\frac{\dot{\phi}}{\phi}\right)^{2}-3\frac{\dot{a}\dot{\phi}}{a\phi}
\end{equation}
\begin{equation}
\label{eff-2}
-2\frac{\ddot{a}}{a}-\frac{\dot{a}^{2}}{a^{2}}  = \frac{8\pi}{\phi c^{2}}p_{eff}+\frac{\omega\dot{\phi}^{2}}{2\phi^{2}}+2\frac{\dot{a}\dot{\phi}}{a\phi}+\frac{\ddot{\phi}}{\phi}
\end{equation} 
For analysis of the universe in derived model, it is convenient to consider equation of state parameter $(\omega_{eff})$ as
$$\omega_{eff} = \frac{p_{eff}}{\rho_{eff}} = \frac{p+\frac{3\phi c^{2}}{32 \pi}\beta^{2}}{\rho+\frac{3\phi c^{2}}{32 \pi}\beta^{2}}$$
For perfect fluid, $p = \gamma \rho$, we obtain $\omega_{eff} \neq -1$. Also, in absence of matter $i. e.\; \rho = p =0$, the effective equation of state parameter is $\omega_{eff} = +1$. That is why the displacement vector $\beta$ can never play the role of a cosmological constant in Brans-Dicke theory for which the equation of state parameter $\omega_{eff} = -1$ is required.\\

From equation (\ref{eff-2}), we obtain the expression for $q$ as following
\begin{equation}
\label{q-1}
q = 1+\frac{8\pi p}{\phi c^{2}H^{2}}+3\frac{\beta^{2}}{4H^{2}}+2\frac{\dot{\phi}}{\phi H}+\frac{\ddot{\phi}}{\phi H^{2}}+\frac{\omega}{2}\frac{\dot{\phi}^{2}}{\phi^{2}H^{2}}
\end{equation}
For acceleration in the derived model, $q < 0$. This means that $1+\frac{8\pi p}{\phi c^{2}H^{2}}+3\frac{\beta^{2}}{4H^{2}}+2\frac{\dot{\phi}}{\phi H}+\frac{\ddot{\phi}}{\phi H^{2}}+\frac{\omega}{2}\frac{\dot{\phi}^{2}}{\phi^{2}H^{2}} < 0$. Since the fluid under taken is barotropic perfect fluid therefore $p > 0$. Also for the model under consideration, $\beta^{2} > 0$. Hence, in derived model, the scalar field $\phi$ and $\omega$ 
contribute acceleration.\\
\section{The model and observational confrontation}\label{3}
\begin{figure}[ht]
\centering
\includegraphics[width=7.5cm,height=6.5cm,angle=0]{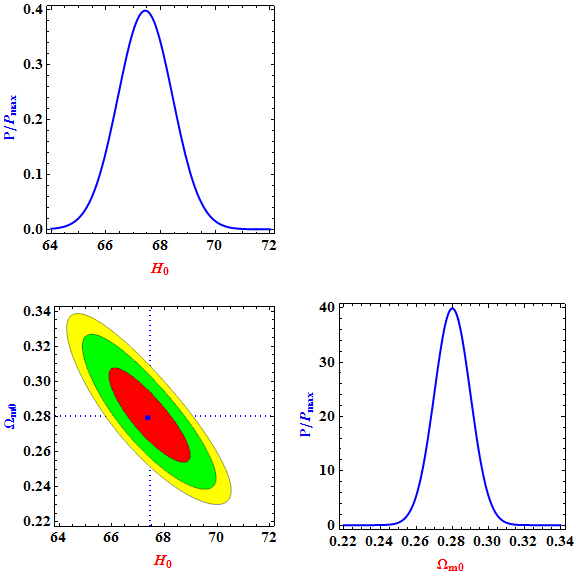} 
\caption{One-dimensional marginalized distributions and two dimensional contours at 1$\sigma$, 2$\sigma $ and 3$\sigma$ confidence regions in $H_{0} - \Omega_{m0}$ plane by bounding our model with 46 OHD points. The estimated values of $H_{0} = 67.44~km~s^{-1}~Mpc^{-1}$ and $\Omega_{m0} = 0.28$. }
\end{figure}
\begin{figure}[ht]
\centering
\includegraphics[width=7.5cm,height=6.5cm,angle=0]{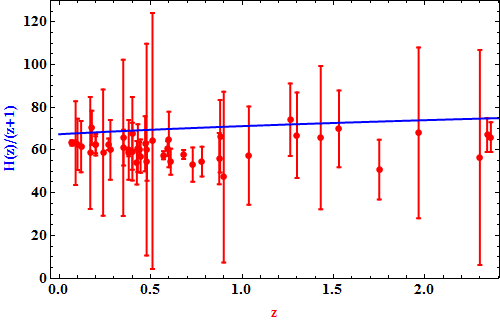}
\caption{The Hubble rate versus red-shift error bar plot with 46 OHD points. The solid line represents our derived model.}
\end{figure}
Since, the present universe is filled with dust or dark matter which pressure is null $(p  = 0)$. Therefore, solving equations (\ref{fe-1}) - (\ref{fe-3}), we obtain
\begin{equation}
\label{m-1}
\frac{\ddot{a}}{a}+2\frac{\dot{a}^{2}}{a^{2}}=(\omega+1)\frac{\ddot{\phi}}{\phi}+(3\omega+2)\frac{\dot{a}\dot{\phi}}{a\phi}
\end{equation}
The general solution of equation (\ref{m-1}) is given by
\begin{equation}
\label{m-2}
\phi = \phi_{0}\left(\frac{a}{a_{0}}\right)^{\frac{1}{\omega+1}}
\end{equation}
where $\phi_{0}$ and $a_{0}$ are constants.\\

Thus, the density parameters are read as
\begin{equation}
\label{dp-1}
\Omega_{m} = \frac{8\pi \rho_{m}}{3c^2 H^2 \phi},~~\Omega_{\beta}=\frac{\beta^{2}}{4H^2} 
\end{equation}
where $\rho_{m} = \left(\rho_{m}\right)_{0}a^{-3}$ and $\Omega_{m}$ and $\Omega_{\beta}$ represent the dimensionless density parameters for matter and $\beta$- energy respectively.\\
Using equation (\ref{dp-1}) in equation (\ref{fe-2}), we obtain
\begin{equation}
\label{m-1}
\Omega_{m}+\Omega_{\beta} = 1+\frac{5\omega+6}{6(\omega+1)^{2}}
\end{equation}
The scale factor in term of red-shift is given by
\begin{equation}
\label{redshift}
a =\frac{a_{0}}{1+z},
\end{equation}
Using equations (\ref{m-2}), (\ref{dp-1}), (\ref{m-1}) and (\ref{redshift}) in equation (\ref{fe-1}), we obtain the  expression for Hubble's function in terms of red-shift as
\begin{widetext}
\begin{equation}
\label{Hubble}
H(z) = \frac{H_{0}}{\sqrt{1+\frac{5\omega+6}{6(\omega+1)^{2}}}}\left[1+\Omega_{m0}(1+z)^{\frac{3\omega+4}{\omega+1}}+\frac{5\omega+6}{6(\omega+1)^{2}}-\Omega_{m0}\right]^{\frac{1}{2}}
\end{equation} 
\end{widetext}
where $H_{0}$ and $\Omega_{m0}$ denote the present values of Hubble constant and matter energy density parameter respectively. In Table \ref{tbl-1}, H(z) is in unit of $km~s^{-1}~Mpc^{-1}$.\\
\begin{figure}[ht]
\centering
\includegraphics[width=7.5cm,height=6.5cm,angle=0]{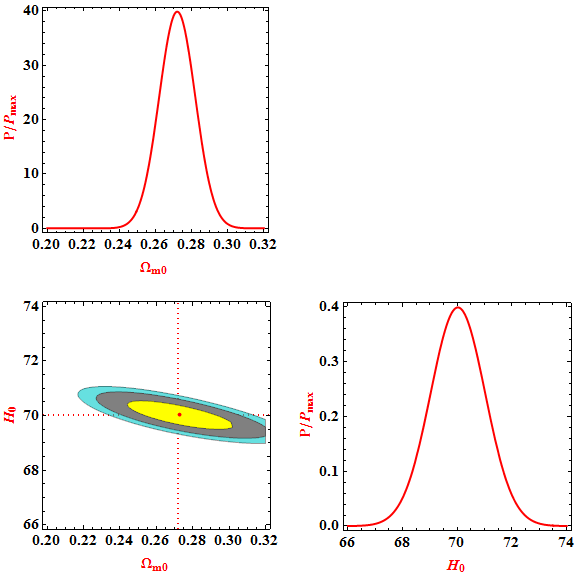} 
\caption{One-dimensional marginalized distributions and two dimensional contours at 1$\sigma$, 2$\sigma $ and 3$\sigma$ confidence regions in $\Omega_{m0} - H_{0}$ plane by bounding our model with SN Ia data points. The estimated values of $H_{0} = 70.02~km~s^{-1}~Mpc^{-1}$ and $\Omega_{m0} = 0.272$.}
\end{figure}
\begin{table*}
\begin{center}
\small
\caption{Hubble parameter H(z) with redshift and errors $\sigma_{i}$.\label{tbl-1}}
\begin{tabular}{@{}crrrrrrrrrrr@{}}
\hline
S.N.~~~~&~~~~z~~~~~& ~~~~H(z)~~~~ & $\sigma_{i}$~~~~ & References \\
\hline
1~~~~&~~~~0~~~~~ &~~~ 67.77~~~ & ~~~ 1.30~~~~ & ~~~~~ \cite{Macaulay/2018} \\
2~~~~&~~~~~0.07~~~~ &~~~~~ 69~~~~ & ~~~~~ 19.6~~~~ & ~~~~~ \cite{Zhang/2014}\\
3~~~~&~~~~~0.09~~~~ &~~~~~ 69~~~~ & ~~~~~ 12~~~~ & ~~~~~ \cite{Simon/2005} \\
4~~~~&~~~~~0.01~~~~ &~~~~~ 69~~~~ & ~~~~~ 12~~~~ & ~~~~~ \cite{Stern/2010} \\
5~~~~&~~~~~0.12~~~~ &~~~~~ 68.6~~~~ & ~~~~~ 26.2~~~~ & ~~~~~ \cite{Zhang/2014}\\
6~~~~&~~~~~0.17~~~~ &~~~~~ 83~~~~ & ~~~~~ 8~~~~ & ~~~~~ \cite{Stern/2010} \\
7~~~~&~~~~~0.179~~~~ &~~~~~75~~~~ & ~~~~~ 4~~~~ & ~~~~~ \cite{Moresco/2012} \\
8~~~~&~~~~~0.1993~~~~ &~~~~~75~~~~ & ~~~~~ 5~~~~ & ~~~~~ \cite{Moresco/2012} \\
9~~~~&~~~~~0.2~~~~ &~~~~~ 72.9~~~~ & ~~~~~ 29.6~~~~ & ~~~~~ \cite{Zhang/2014}\\
10~~~~&~~~~~0.24~~~~ &~~~~~ 79.7~~~~ & ~~~~~ 2.7~~~~ & ~~~~~ \cite{Gazta/2009}\\
11~~~~&~~~~~0.27~~~~ &~~~~~ 77~~~~ & ~~~~~ 14~~~~ & ~~~~~ \cite{Stern/2010} \\
12~~~~&~~~~~0.28~~~~ &~~~~~ 88.8~~~~ & ~~~~~ 36.6~~~~ & ~~~~~ \cite{Zhang/2014}\\
13~~~~&~~~~~0.35~~~~ &~~~~~ 82.7~~~~ & ~~~~~ 8.4~~~~ & ~~~~~ \cite{Chuang/2013}\\
14~~~~&~~~~~0.352~~~~ &~~~~~83~~~~ & ~~~~~ 14~~~~ & ~~~~~ \cite{Moresco/2012} \\
15~~~~&~~~~~0.38~~~~ &~~~~~81.5~~~~ & ~~~~~ 1.9~~~~ & ~~~~~ \cite{Alam/2016} \\
16~~~~&~~~~~0.3802~~~~ &~~~~~83~~~~ & ~~~~~ 13.5~~~~ & ~~~~~ \cite{Moresco/2016} \\
17~~~~&~~~~~0.4~~~~ &~~~~~ 95~~~~ & ~~~~~ 17~~~~ & ~~~~~ \cite{Simon/2005} \\
18~~~~&~~~~~0.4004~~~~ &~~~~~77~~~~ & ~~~~~ 10.2~~~~ & ~~~~~ \cite{Moresco/2016} \\
19~~~~&~~~~~0.4247~~~~ &~~~~~87.1~~~~ & ~~~~~ 11.2~~~~ & ~~~~~ \cite{Moresco/2016} \\
20~~~~&~~~~~0.43~~~~ &~~~~~ 86.5~~~~ & ~~~~~ 3.7~~~~ & ~~~~~ \cite{Gazta/2009}\\
21~~~~&~~~~~0.44~~~~ &~~~~~ 82.6~~~~ & ~~~~~ 7.8~~~~ & ~~~~~ \cite{Blake/2012}\\
22~~~~&~~~~~0.44497~~~~ &~~~~~ 92.8~~~~ & ~~~~~ 12.9~~~~ & ~~~~~ \cite{Moresco/2016}\\
23~~~~&~~~~~0.47~~~~ &~~~~~ 89~~~~ & ~~~~~ 49.6~~~~ & ~~~~~ \cite{Ratsimbazafy/2017}\\
24~~~~&~~~~~0.4783~~~~ &~~~~~80.9~~~~ & ~~~~~ 9~~~~ & ~~~~~ \cite{Moresco/2016} \\
25~~~~&~~~~~0.48~~~~ &~~~~~ 97~~~~ & ~~~~~ 60~~~~ & ~~~~~ \cite{Stern/2010} \\
26~~~~&~~~~~0.51~~~~ &~~~~~90.4~~~~ & ~~~~~ 1.9~~~~ & ~~~~~ \cite{Alam/2016} \\
27~~~~&~~~~~0.57~~~~ &~~~~~96.8~~~~ & ~~~~~ 3.4~~~~ & ~~~~~ \cite{Anderson/2014} \\
28~~~~&~~~~~0.593~~~~ &~~~~~104~~~~ & ~~~~~ 13~~~~ & ~~~~~ \cite{Moresco/2012} \\
29~~~~&~~~~~0.6~~~~ &~~~~~ 87.9~~~~ & ~~~~~ 6.1~~~~ & ~~~~~ \cite{Blake/2012}\\
30~~~~&~~~~~0.61~~~~ &~~~~~97.3~~~~ & ~~~~~ 2.1~~~~ & ~~~~~ \cite{Alam/2016} \\
31~~~~&~~~~~0.68~~~~ &~~~~~92~~~~ & ~~~~~ 8~~~~ & ~~~~~ \cite{Moresco/2012} \\
32~~~~&~~~~~0.73~~~~ &~~~~~ 97.3~~~~ & ~~~~~ 7~~~~ & ~~~~~ \cite{Blake/2012}\\
33~~~~&~~~~~0.781~~~~ &~~~~~105~~~~ & ~~~~~ 12~~~~ & ~~~~~ \cite{Moresco/2012} \\
34~~~~&~~~~~0.875~~~~ &~~~~~125~~~~ & ~~~~~ 17~~~~ & ~~~~~ \cite{Moresco/2012} \\
35~~~~&~~~~~0.88~~~~ &~~~~~ 90~~~~ & ~~~~~ 40~~~~ & ~~~~~ \cite{Stern/2010} \\
36~~~~&~~~~~0.9~~~~ &~~~~~ 117~~~~ & ~~~~~ 23~~~~ & ~~~~~ \cite{Stern/2010} \\
37~~~~&~~~~~1.037~~~~ &~~~~~154~~~~ & ~~~~~ 20~~~~ & ~~~~~ \cite{Moresco/2012} \\
38~~~~&~~~~~1.3~~~~ &~~~~~ 168~~~~ & ~~~~~ 17~~~~ & ~~~~~ \cite{Stern/2010} \\
39~~~~&~~~~~1.363~~~~ &~~~~~160~~~~ & ~~~~~ 33.6~~~~ & ~~~~~ \cite{Moresco/2015} \\
40~~~~&~~~~~1.43~~~~ &~~~~~ 177~~~~ & ~~~~~ 18~~~~ & ~~~~~ \cite{Stern/2010} \\
41~~~~&~~~~~1.53~~~~ &~~~~~ 140~~~~ & ~~~~~ 14~~~~ & ~~~~~ \cite{Stern/2010} \\
42~~~~&~~~~~1.75~~~~ &~~~~~ 202~~~~ & ~~~~~ 40~~~~ & ~~~~~ \cite{Stern/2010} \\
43~~~~&~~~~~1.965~~~~ &~~~~~186.5~~~~ & ~~~~~ 50.4~~~~ & ~~~~~ \cite{Moresco/2015} \\
44~~~~&~~~~~2.3~~~~ &~~~~~ 224~~~~ & ~~~~~ 8~~~~ & ~~~~~ \cite{Busca/2013} \\
45~~~~&~~~~~2.34~~~~ &~~~~~ 222~~~~ & ~~~~~ 7~~~~ & ~~~~~ \cite{Delubac/2015} \\
46~~~~&~~~~~2.36~~~~ &~~~~~226~~~~ & ~~~~~ 8~~~~ & ~~~~~ \cite{Ribera/2014} \\
\hline
\end{tabular}
\end{center}
\end{table*}
\begin{itemize}
\item {\bf Observational Hubble Data (OHD)}: We adopt $46$ Observational Hubble Data (OHD) points over the redshift range of $0\leq z\leq 2.36$ (see table \ref{tbl-1}), obtained from cosmic chronometric (CC) technique. Since CC technique is based on the \textquotedblleft galaxy differential age\textquotedblright method therefore OHD data obtained from CC technique is model-independent and not correlated.\\

\item {\bf SN Ia Data}: We use SN Ia data which includes 9 high $z$ and 27 low $z$, Gold Sample and New Gold Sample \cite{Vishwakarma/2018} and references therein.\\ 
\end{itemize}
For statistical analysis, we define $\chi^{2}$ as following:
\begin{equation}
\label{chi1}
\chi^{2}_{OHD} = \sum_{i=1}^{46}\left[\frac{H_{th}(z_{i})-H_{obs}(z_{i})}{\sigma_{i}}\right]^{2}
\end{equation}
where $H_{obs}(z_{i})$ is the observed value of Hubble parameter with standard deviation $\sigma_{i}$ and 
$H_{th}(z_{i})$ is the theoretical value obtained from bounding equation (\ref{chi1}) with 46 OHD points. The estimated values of $H_{0} = 67.44~km~s^{-1}~Mpc^{-1}$ and $\Omega_{m0} = 0.28$ with $\chi^{2}_{min} = 26.36$.\\

Similarly for SN Ia data, we have estimated the model parameters $H_{0}$ and $\Omega_{m0}$ as $H_{0} = 70.02~km~s^{-1}~Mpc^{-1}$ and $\Omega_{m0} = 0.272$ with $\chi^{2}_{min} = 562.22$. The numerical result is summarized in Table II. Figures 1 and 3 depict one-dimensional marginalized distributions and two dimensional contours at 1$\sigma$, 2$\sigma $ and 3$\sigma$ confidence regions by bounding our model with OHD and SN Ia data points respectively. The best fit curve for Hubble rate is exhibited in Figure 2.
\begin{table*}
\begin{center}
\small
\caption{Summary of the numerical result.\label{tbl-2}}
\begin{tabular}{@{}crrrrrrrrrrr@{}}
\hline
Model parameters ~~~~~ & ~~~~~ OHD ~~~~~ & ~~~~~ SN Ia \\
\hline
$\Omega_{m0}$ ~~~~~ &~~~~~$0.28$ ~~~~~ & ~~~~~ $0.272$  \\
$H_{0}$ ~~~~~ & ~~~~~ $67.44$ ~~~~~ & ~~~~~ $70.02$ \\
\hline
\end{tabular}
\end{center}
\end{table*}
\section{Cosmological parameters of the model}\label{4}
\subsection{Deceleration parameter}
The deceleration parameter of derived model in terms of red-shift is read as
\begin{equation}
\label{dp}
q=-1+\frac{(1+z)H^{\prime}(z)}{H(z)}
\end{equation}
where $H^{\prime}(z)$ is the first order derivative of H(z) with respect to z.
\begin{figure}[ht]
\centering
\includegraphics[width=7.5cm,height=6.0cm,angle=0]{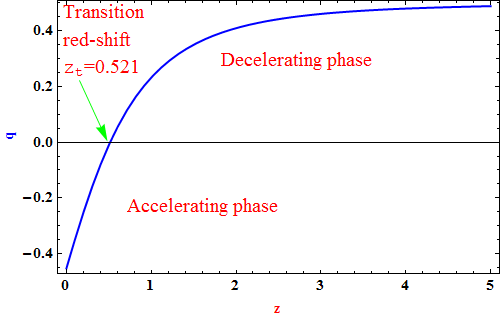}
\caption{The variation of $q$ versus $z$ for $\Omega_{m0} = 0.28$, $H_{0} = 67.44~km~s^{-1}~Mpc^{-1}$ and $\omega = 40000$.}
\end{figure}
Solving equations (\ref{Hubble}) and (\ref{dp}), we get
\begin{widetext}
\begin{equation}
\label{dp-1}
q = -1 + \frac{(3 \omega +4) \Omega _{m0} (z+1)^{\frac{3 \omega +4}{\omega +1}}}{2 (\omega +1) \left(-\Omega _{m0}+\Omega _{m0} (z+1)^{\frac{3 \omega +4}{\omega +1}}+\frac{5 \omega +6}{6 (\omega +1)^2}+1\right)}
\end{equation}
\end{widetext}
The graphical behaviour of deceleration parameter is depicted in Figure 4. From equation (\ref{dp-1}), one can obtain the present value of deceleration parameter as $q_{0} =  -0.58$ by putting $z = 0$, $\Omega_{m0} = 0.28$ and $\omega = 40000$. This value of $q_{0}$ is in pretty agreement with recent observations.\\

Figure 4 shows the signature flipping behavior of deceleration parameter with decreasing value of redshift i.e. at beginning $q$ was evolving with positive sign which indicates that early universe was in decelerating phase and it turns into accelerating mode at $z_{t} = 0.521$. This transitioning evolution of $q$ from positive to negative value leads the concept of hybrid universe. In recent past, various cosmological models with hybrid expansion law have been investigated in different physical contexts \cite{Yadav/2019,Akarsu/2014,Yadav/2019b,Yadav/2012,Pradhan/2011,Mishra/2018,Ray/2019,Yadav/2015,Yadav/2016}.
\subsection{Age of the universe}\label{5} 
\begin{figure}[ht]
\centering
\includegraphics[width=7.5cm,height=6.5cm,angle=0]{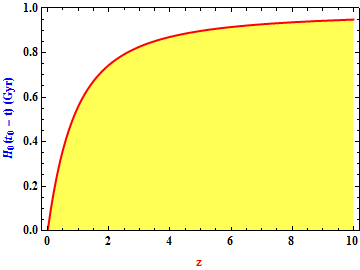}
\caption{The plot of $H_{0}(t_{0}-t)$ versus $z$ for $\Omega_{m0} = 0.28$, $H_{0} = 67.44~km~s^{-1}~Mpc^{-1} \sim 0.0688\;Gyrs$ and $\omega = 40000$. Here, we observe that for $z = 0$, $H_{0}(t_{0} - t)$ becomes null which turns into imply that at present $t = t_{0}$.}
\end{figure}
The age of scalar field Brans-Dike universe is obtained as 
\begin{equation}
\label{age-1}
H(z) = -\frac{1}{1+z}\frac{dz}{dt}\;\; \Rightarrow\;\; dt = -\frac{dz}{(1+z)H(z)}
\end{equation}
Hence,
\begin{widetext}
\begin{equation}
\label{age-2}
\int_{t_{0}}^{t}dt = -\int_{z}^{0}\frac{dz}{(1+z)\frac{H_{0}}{\sqrt{1+\frac{5\omega+6}{6(\omega+1)^{2}}}}\left[1+\Omega_{m0}(1+z)^{\frac{3\omega+4}{\omega+1}}+\frac{5\omega+6}{6(\omega+1)^{2}}-\Omega_{m0}\right]^{\frac{1}{2}}}
\end{equation}
\end{widetext}
Here, $t_{0}$ is the present age of the universe and it is obtained as
\begin{widetext}
\begin{equation}
\label{age-3}
t_{0} = \lim_{x\rightarrow\infty}{\int_{0}^{z}\frac{dz}{(1+z)\frac{H_{0}}{\sqrt{1+\frac{5\omega+6}{6(\omega+1)^{2}}}}\left[1+\Omega_{m0}(1+z)^{\frac{3\omega+4}{\omega+1}}+\frac{5\omega+6}{6(\omega+1)^{2}}-\Omega_{m0}\right]^{\frac{1}{2}}}}
\end{equation}
\end{widetext}
Integrating equation (\ref{age-3}), we obtain
\begin{equation}
\label{age-4}
H_{0}t_{0} = 0.983019
\end{equation}

Thus the present age of the universe is computed as $t_{0} = 0.983019~H_{0}^{-1} = 14.288\;Gyrs$. It is important to note that the present age of the universe in derived model nicely match with its empirical value, predicted by recent WMAP observations \cite{Hinshaw/2013} and Plank collaborations \cite{Ade/2014}. Therefore, the model under consideration have pretty consistency with recent astrophysical observations \cite{Ade/2014,Hinshaw/2013}.\\  
\subsection{Jerk parameter}
The jerk parameter (j) \cite{Mukherjee/2019}, in terms of red-shift and Hubble function is read as
\begin{equation}
\label{jerk-1}
1-(1+z)\frac{[H(z)^{2}]^{\prime}}{H(z)^{2}}+\frac{1}{2}(1+z)^{2}\frac{[H(z)^{2}]^{\prime\prime}}{H(z)^{2}}
\end{equation}
Equations (\ref{Hubble}) and (\ref{jerk-1}) lead to
\begin{widetext}
\[
j=1-\frac{3 (\omega +1) (3 \omega +4) \Omega _{m0} (z+1)^{\frac{3 \omega +4}{\omega +1}}}{6 (\omega +1)^2 \Omega _{m0} \left((z+1)^{\frac{3 \omega +4}{\omega +1}}-1\right)+6 \omega ^2+17 \omega +12}+
\]  
\begin{equation}
\label{jerk-2}
\frac{9 (3 \omega +4)^2 \Omega _{m0}^2 (z+1)^{\frac{4 \omega +6}{\omega +1}} \left(3 (\omega +1)^2 \Omega _{m0} \left(2 \left((z+1)^{\frac{3 \omega +4}{\omega +1}}-3\right)+\omega  \left((z+1)^{\frac{3 \omega +4}{\omega +1}}-4\right)\right)+\xi\right)^{2}}{2 \left(6 (\omega +1)^2 \Omega _{m0} \left((z+1)^{\frac{3 \omega +4}{\omega +1}}-1\right)+6 \omega ^2+17 \omega +12\right)^{4}}
\end{equation}
\end{widetext}
where $\xi = (2 \omega +3)^2 (3 \omega +4)$.\\

The explicit expression of jerk parameter is given in equation (\ref{jerk-2}) and its behavior is graphed in Figure 5. We observe that for high red-shift values, the jerk parameter have its low values and it increases as $z$ decreases. The present value of jerk parameter is computed as $j_{0} = 0.8002$. Therefore, the derived model represents a model of the universe other than $\Lambda$CDM universe. Note that $H$, $q$ and $j$ are the important parameters of cosmographic series. The usefulness of these parameters in discriminating the various dark energy models are given in Refs. \cite{CAPOZZIELLO/2019,Dunsby/2016}.\\ 
\begin{figure}[ht]
\centering
\includegraphics[width=7.5cm,height=6.5cm,angle=0]{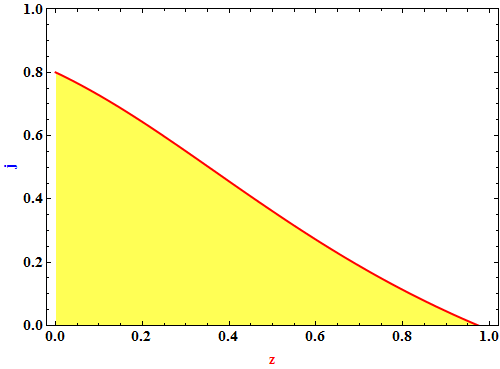}
\caption{The plot of $j$ versus $z$ for $\Omega_{m0} = 0.28$ and $\omega = 40000$.}
\end{figure}
\section{Concluding remarks}\label{7}
In this paper, we have investigated a scalar field Brans-Dicke cosmological model in Lyra's manifold and also checked its validity by performing well known $\chi^{2}$ test for free parameters of the model with recent observational H(z) and SN Ia data. The derived model successfully passes this test on the scale of statistical analysis and represents the best fit curve for Hubble rate (see Fig. 2). The present values of deceleration parameter and age of the universe are calculated in section IV. We observe that the model under consideration have pretty consistency with recent Plank collaboration results and WMAP observations which in turn implies that the derived model is physically viable. Figure 4 exhibits that the universe in derived model evolves with negative deceleration parameter i.e. the present universe expands with acceleration due to accumulation of matter in scalar field. So, the proposed model describes features of the universe from early decelerating phase to current accelerating phase without inclusion of any exotic type matter or energy. The natural behavior of jerk parameter of scalar field Brans-Dicke universe in Lyra's manifold is shown in Figure 6. It is worthwhile to note that the displacement vector $\beta$ does not behave like $\Lambda$ - particularly in contributing the late time acceleration of the universe however the scalar field $\phi$ dominates the current universe. Finally in spite of very good possibility of co-existence of Brans-Dicke gravity and Lyra's geometry to provide a theoretical foundation for relativistic gravitation, astrophysics and cosmology, the experimental point is yet to be considered. But still the theory needs a fair trial.\\

\textbf{Acknowledgments:} The authors are grateful to the reviewer for illuminating suggestions that have significantly improved our work in terms of research quality and presentation. The authors are grateful to Y. Heydarzade for fruitful comments on the paper.


\end{document}